\documentstyle[emulateapj,epsf]{article}

\lefthead{Larkin et al.}
\righthead{Obscured Broad Line in MRC2025}
\begin{document}
  
\title{Discovery of an Obscured Broad Line Region in the High Redshift
Radio Galaxy MRC~2025-218}

\author{James E. Larkin$^1$, Ian S. McLean$^1$, James R. Graham$^2$,
E. E. Becklin$^1$, Donald F.  Figer$^3$, Andrea M. Gilbert$^2$,
N. A. Levenson$^{2,4}$, Harry I. Teplitz$^{5,6}$,
Mavourneen K. Wilcox$^1$ \& Tiffany M. Glassman$^1$  }

\affil{$^1$Dept. of Physics and Astronomy, University of California,
Los Angeles,
$^2$Dept. of Astronomy, University of California, Berkeley,
$^3$Space Telescope Science Institute,
$^4$Dept. of Physics and Astronomy, John's Hopkins University,
$^5$LASP, Goddard Space Flight Center,
$^6$NOAO Research Associate}
\authoremail{larkin@astro.ucla.edu}

\begin{abstract}
This paper presents infrared spectra taken with the newly commissioned
NIRSPEC spectrograph on the Keck II Telescope of the High Redshift
Radio Galaxy MRC~2025-218 (z=2.63) These observations represent the
deepest infrared spectra of a radio galaxy to date and have allowed
for the detection of H$\beta$, [OIII] (4959/5007), [OI] (6300), H$\alpha$,
[NII] (6548/6583) and [SII] (6716/6713).  The H$\alpha$ emission is very
broad (FWHM = 9300 km/s) and luminous (2.6$\times$10$^{44}$ ergs/s)
and it is very comparable to the line widths and strengths of radio
loud quasars at the same redshift. This strongly supports AGN
unification models linking radio galaxies and quasars, although we
discuss some of the outstanding differences.  The [OIII] (5007) line is
extremely strong and has extended emission with large relative
velocities to the nucleus.  We also derive that if the extended
emission is due to star formation, each knot has a star formation rate
comparable to a Lyman Break Galaxy at the same redshift.

\end{abstract}


\section{Introduction}

Deep radio surveys have proven to be one of the best methods for
finding high redshift galaxies. Most evidence suggests that these
powerful radio sources are the precursors of local giant ellipticals
(e.g. Pentericci, et al. 1999).  Many have irregular and complex
morphologies suggestive of mergers and they are often surrounded by an
overdensity of compact sources; presumably sub-galactic clumps
(e.g. van Breugel et al. 1998).  At both low and high redshifts, radio
galaxies usually have strong optical emission lines, especially OIII
at 5007~\AA.  It is strongly debated, however, if the emission lines
arise by the same mechanism as the radio jets. Several authors
(e.g. Rawlings \& Saunders 1992, Eales \& Rawlings 1993, and Evans
1998) have demonstated a strong correlation between radio luminosity
and [OIII] luminosity, but as Evans showed, there is a strong selection
effect based on the detection limits as a function of distance and
this may explain much of the correlation.  Since the galaxies are
often disturbed, star formation, large scale shocks and a central AGN
are all possible sources of the line emission.  Active galaxy
unification models suggest that radio galaxies are quasars with
obscured broad line regions (e.g. Antonucci 1993). Eales \& Rawlings
(1993, 1996) and Evans (1998) have been successful at using infrared
spectrographs on 4-meter class telescopes to measure a few of the
brightest lines in small samples of radio galaxies in the redshift
range 2.2 to 2.6 and they find line ratios most consistent with
Seyfert 2 (obscured AGN) nuclei.  Independent of our current efforts,
a team has also successfully used the ISAAC instrument on the VLT to
observe high redshift radio galaxies (HzRGs) including
MRC~2025-218 (McCarthy, personal communication).

An additional unexplained phenomenon is that at high redshifts
(z$>$0.6) the radio, optical continuum, infrared continuum and
emission line structures tend to be closely aligned (Chambers, Miley,
\& van Bruegel 1987, and McCarthy et al. 1987). This is probably not
seen in lower redshift targets because the central activity tends to
be a smaller fraction of the total luminosity than in high redshift
sources. Among the proposed explanations are that the emission
lines arise from shock induced star formation (De Young 1989, Rees
1989) or that it is scattered light originating from the central
nucleus (Fabian 1991).  This is a crucial question in our
understanding of how and when most star formation occured in giant
elliptical galaxies and in clusters in general.

MRC~2025-218 (z=2.630) has a compact near infrared and optical continuum
morphology (van Breugel et al. 1998), but extended Ly$\alpha$ emission
(5$^{\prime \prime}$) aligned with its radio axis (McCarthy et
al. 1992). The extended UV emission has significant (8.3$\pm$2.3 \%)
linear polarization perpendicular to the UV axis (Cimatti et al. 1996)
suggesting that scattering plays a significant role.  McCarthy et
al. also found three extremely red galaxies (ERO's: R-K $>$ 6 mag)
within 20$^{\prime \prime}$ of the radio galaxy.  This is a large
overdensity of such objects and strongly suggests an association
between the ERO's and the active galaxy.  In this paper we present
infrared spectra taken with a long slit oriented close to the radio
axis and including one of the ERO's.  The ERO spectra will be
described in a future paper. For all calculations we have assumed a
cosmology with $\Lambda$=0, q$_0$=0.1 and H$_0$=75 km s$^{-1}$
Mpc$^{-1}$.  For a redshift of 2.63 this yields a luminosity distance
of 2.1$\times$10$^4$ Mpc and an angular scale of 7.7 kpc per
arcsecond.

\section{Observations and Data Reduction}

The field of MRC~2025-218 was observed on 4 Jun, 1999 (UT) with the
near infrared spectrograph NIRSPEC (McLean, et al. 1998 and McLean, et
al.  2000) on the Keck II Telescope during its commissioning.  First
the field was imaged in the K-band with the slitviewing camera which
is a HgCdTe PICNIC detector (256$^2$ pixels) sensitive from 1 to 2.5
microns. Figure 1 shows the reduced image of the field with a total
integration time of 540 seconds and a FWHM of 0\farcs54. As shown in
the figure, the slit (42$^{\prime \prime}$ long and 0\farcs57 wide)
was placed on both the radio galaxy and the extremely red galaxy
dubbed ERO-A by McCarthy et al. (1992).  This corresponded to a slit
position angle of -7 degrees.

{\plotfiddle{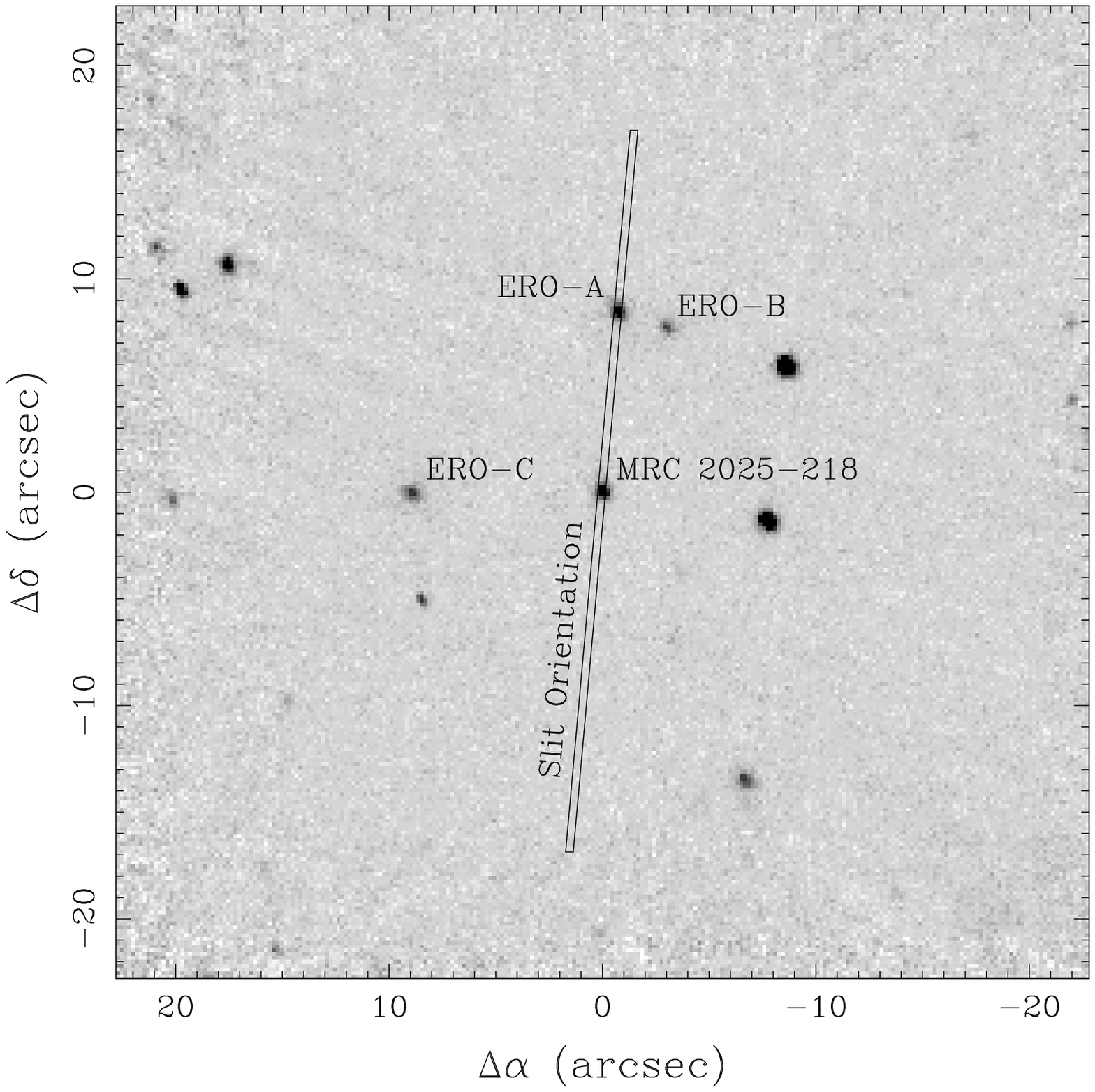}{2.8in}{0}{40}{40}{0}{-55}}
{\footnotesize Figure 1. - K band image of the MRC 2025-218 field.}
\medskip

For spectroscopy, the telescope was repeatedly moved roughly 20
arcseconds to center the objects first in the upper portion of the
slit then the lower portion.  Four 300 second exposures were taken in
both the H-band ($\sim$1.6$\mu$m) and K-band ($\sim$2.2$\mu$m)
yielding an effective integration time on MRC~2025-215 of 20 minutes
in each band.  For guiding, NIRSPEC's optical guide camera was used to
actively track a bright star roughly 2 arcminutes from MRC~2025-218.

Arc lamp and flat lamp spectra were taken at each setting prior to
changing mechanism setups.  The 7.6 magnitude A0 star PPM~272233 was
also observed at the same settings in order to remove telluric
absorption effects from the atmosphere.  The calibrator star was
reduced first. For each band the spectral pair was subtracted and
divided by a reduced flat field lamp spectra.  Bad pixels were then
identified and removed by medianing the four nearest neighbors.  The
spectra were spatially rectified using a quadratic polynomial at each
row, then spectrally rectified with a quadratic at each column.  The
negative spectrum of the star was then shifted and subtracted from the
positive spectrum producing a combined spectrum with residual
atmospheric lines removed.  The stellar spectrum was extracted by
averaging the central 3 pixels along the 2-d spectrum.  A synthetic
black body spectrum was divided into the stellar spectrum and residual
hydrogen absorption lines from the Brackett series were interpolated
over. The spectra of the radio galaxy were reduced in a similar way
except they were divided by the reduced calibration star spectrum
instead of a black body. For extraction of the galaxy spectra, a 6
pixel spatial aperture (1\farcs14) was used. Spectrophotometry was
obtained by determining the equivalent widths of the emission lines
within a 1\farcs5 aperture in the spectra and comparing this to the
broad band fluxes of the galaxy in a 1\farcs5 circular aperture in the
slit viewing camera images.

{\plotfiddle{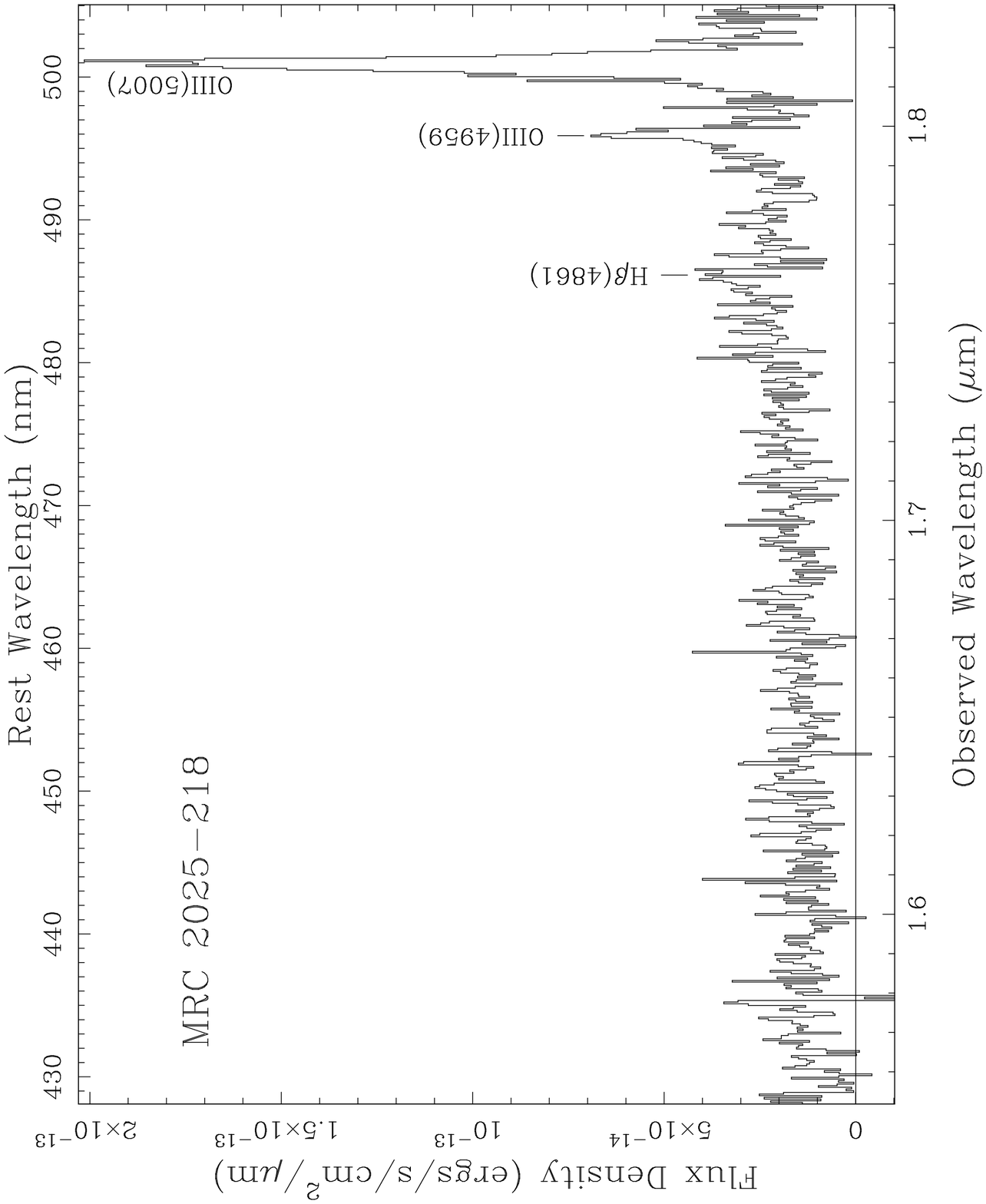}{2.9in}{270}{38}{38}{-15}{230}}
{\footnotesize Figure 2. - H band spectrum of MRC 2025-218. It is
dominated by [OIII] at rest wavelength 5007~\AA. Also present is the
other member of this doublet ([OIII], 4959~\AA) and a weak H$\beta$
emission line.}
\medskip

\section{Results}

Figure 2 shows the H-band spectrum of MRC~2025-218.  By far the most
dominant line is [OIII] (5007 \AA) redshifted to 1.82 $\mu$m.  This line
is highlighted in figure 3 where the complete position velocity map of
this line is presented. Panel (a) of figure 3 is stretched to
highlight the spectrally double nature of the nuclear emission
($\Delta$v $\sim$ 200 km sec$^{-1}$).  Panel (b) shows three faint
emission knots at large angular separations (1$^{\prime
\prime}$-2$^{\prime \prime}$) and/or high kinematic velocities
($\sim$400 km s$^{-1}$) Although faint, these structures repeat in the
individual spectra that cover the [OIII] line.  Two knots appear at
essentially 0 km sec$^{-1}$ relative velocity, but 1\farcs8 North and
2\farcs4 South of the Nucleus.  A high speed clump appears 1$^{\prime
\prime}$ North of the nucleus and at a redshifted relative velocity of
410 km sec$^{-1}$.  This high speed clump is also the brightest within
our slit with a flux of roughly 1$\times$10$^{-16}$ ergs s$^{-1}$
cm$^{-2}$. Also detected in the H-band spectrum is the other member of
the [OIII] doublet at 4959 \AA, and H$\beta$.  The ratio of
[OIII]~/~H$\beta$ is extremely large at 17$\pm$7.  The H$\beta$ line has
a total nuclear flux of only
5~$\times$~10$^{-17}$~ergs~cm$^{-2}$~s$^{-1}$.  Table 1 gives all
detected fluxes and line widths.

Figure 4 shows the K-band spectrum which is dominated by a broad
H$\alpha$ emission line.  The spectrum has had a median filter passed
over it to improve the appearance of the fainter lines.  The H$\alpha$
is well modeled by a pair of Gaussians having line widths of
9300$\pm$900 km/s and 730$\pm$100 km/s.  The narrow component is
consistent with the Ly$\alpha$ line width of 700 km/s found by
Villar-Martin et al. 1999.  After subtracting away the two H$\alpha$
components, the middle graph in figure 4 shows the strong [NII]
(6548/6583 \AA) emission lines as well as weaker features from [OI]
(6300 \AA) and [SII](6716/6731 \AA).  The line fluxes and widths are
also given in table 1.

{\plotfiddle{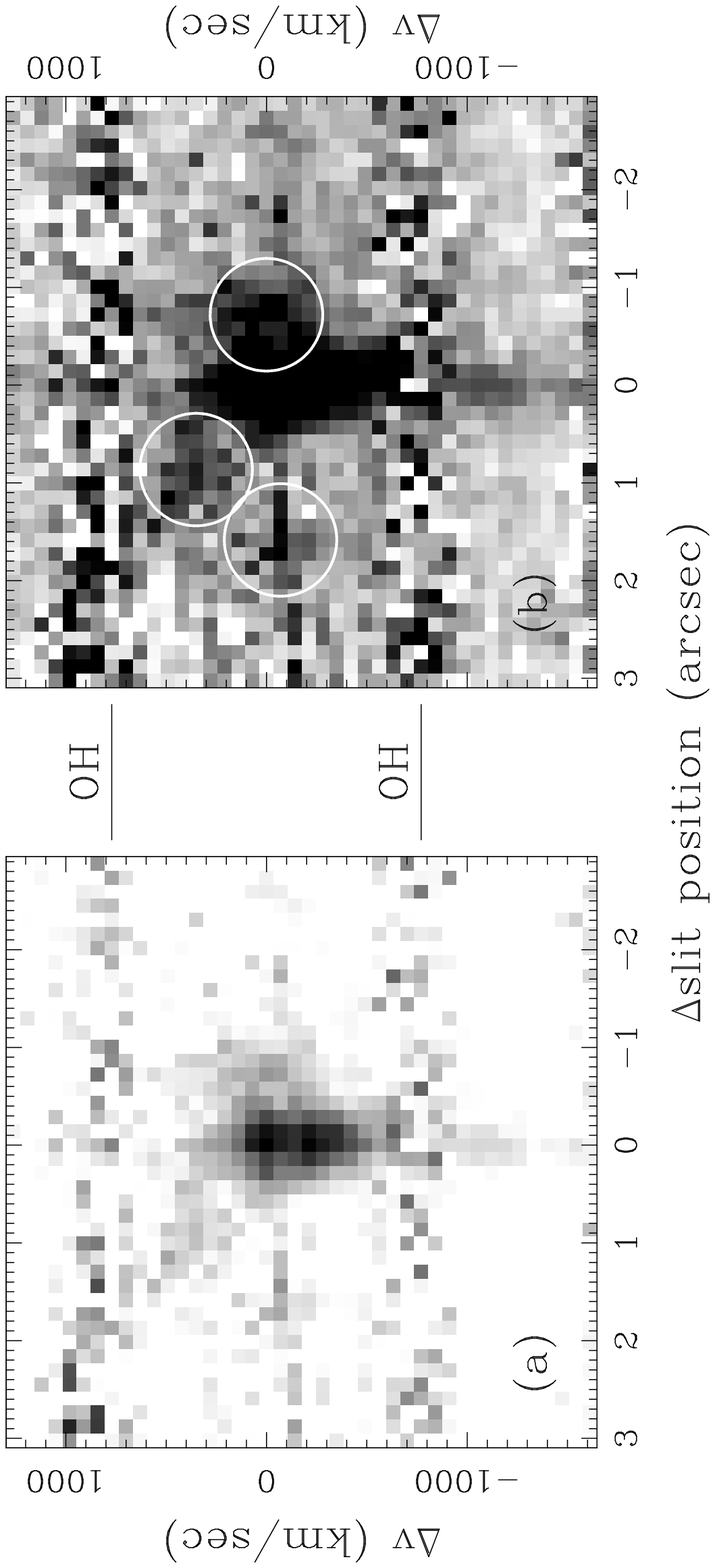}{1.9in}{270}{40}{40}{-30}{190}} {\footnotesize
Figure 3. - Position velocity plots for OIII (5007). Panel (a) is
stretched to show the double nuclear peak.  Panel (b) highlights three
extended emission regions circled in white.  The OIII line is highly
disturbed with several different kinematic and spatial components
including a kinematically split nucleus and a high velocity (400 km/s)
knot located 2'' off nucleus.  Nearby OH lines from the Earth's
atmosphere are labeled.}
\medskip

\smallskip
{\centering
\footnotesize \begin{tabular}{lccc}
\multicolumn{4}{c}{\bf TABLE 1} \\
\multicolumn{4}{c}{\bf Emission Line Strengths} \\
\hline
\hline
\multicolumn{1}{c}{} & Rest & Flux & Line Width \\
\multicolumn{1}{c}{Line} & $\lambda$(\AA) & ($\times 10^{-16}$
ergs s$^{-1}$ cm$^{-2}$) & (km s$^{-1}$) \\
\hline
SII & 6731 & 0.4$\pm$0.3 & 200$\pm$100 \\
SII & 6716 & 0.6$\pm$0.3 & 200$\pm$100 \\
NII & 6583 & 1.3$\pm$0.3 & 880$\pm$100 \\
NII & 6548 & 1.3$\pm$0.3 & 880$\pm$100 \\
H$\alpha$(narrow) & 6563 & 2.7$\pm$0.4 & 730$\pm$100 \\
H$\alpha$(broad) & 6563 & 18$\pm$2 & 9300$\pm$900 \\
OI & 6300 & 0.8$\pm$0.3 & 800$\pm$400 \\
OIII & 5007 & 8.4$\pm$1.6 & 600$\pm$200 \\
OIII & 4959 & 2.1$\pm$0.4 & 600$\pm$200 \\
H$\beta$ & 4861 & 0.5$\pm$0.3 & 600$\pm$200 \\
\hline
\end{tabular} }

\section{Discussion}

\subsection{ Nuclear Spectrum}

The nuclear spectrum of the HzRG MRC~2025-218 is clearly dominated by
emission lines from a central AGN.  The broad H$\alpha$ line width is
9300 km/s which is only seen in type I AGN (unobscured broad line
regions).  This line width is very close to the mean H$\beta$ line
width of 9870$\pm$950 km/s of radio loud quasars in the redshift range
2.0 to 2.5 by McIntosh et al., 1999.  The ratio of
[OIII]/H$\beta$ is 17 which is also only seen in AGN and ratios of
[NII]/H$\alpha$ and [OI]/H$\alpha$ are also consistent with AGN
excitation (Osterbrock, 1989).

From the H$\alpha$/H$\beta$ narrow line ratio of 5.4 we derive an
optical extinction A$_V$=1.4 mag. In this calculation we've assumed an
intrinsic ratio of H$\alpha$/H$\beta$ = 3.1 as seen in local AGN
(Osterbrock, 1989), and the interstellar extinction law of Cardelli et
al. (1989).  This must be treated as an upper limit, however, since
radio loud objects may have elevated H$\alpha$ due to collisional
excitation (e.g. Baker et al. 1994). If the broad line ratio of
H$\alpha$/H$\beta$ were similar to the narrow line ratio, then broad
H$\beta$ should have marginally been detected in our H-band spectrum.
We therefore feel safe in the assumption that the extinction to the
broad line region is similar to the value for the narrow line region
(1.4 mag), but not necessarily significantly greater.  This extinction
is also sufficient to explain the lack of broad Ly$\alpha$ detections
in McCarthy et al. (1990) and Villar-Martin et al. (1999b). Without
extinction our broad line emission would predict a Ly$\alpha$ broad
line flux of 2.6$\times$10$^{-15}$~ergs~s$^{-1}$~cm$^{-2}$ in the
Villar-Martin slit which would have been easily detected but with
A$_V$=1.4 mag this is reduced to less than 3$\times$10$^{-16}$
ergs~s$^{-1}$~cm$^{-2}$ which would have been marginably detected at
best.

{\plotfiddle{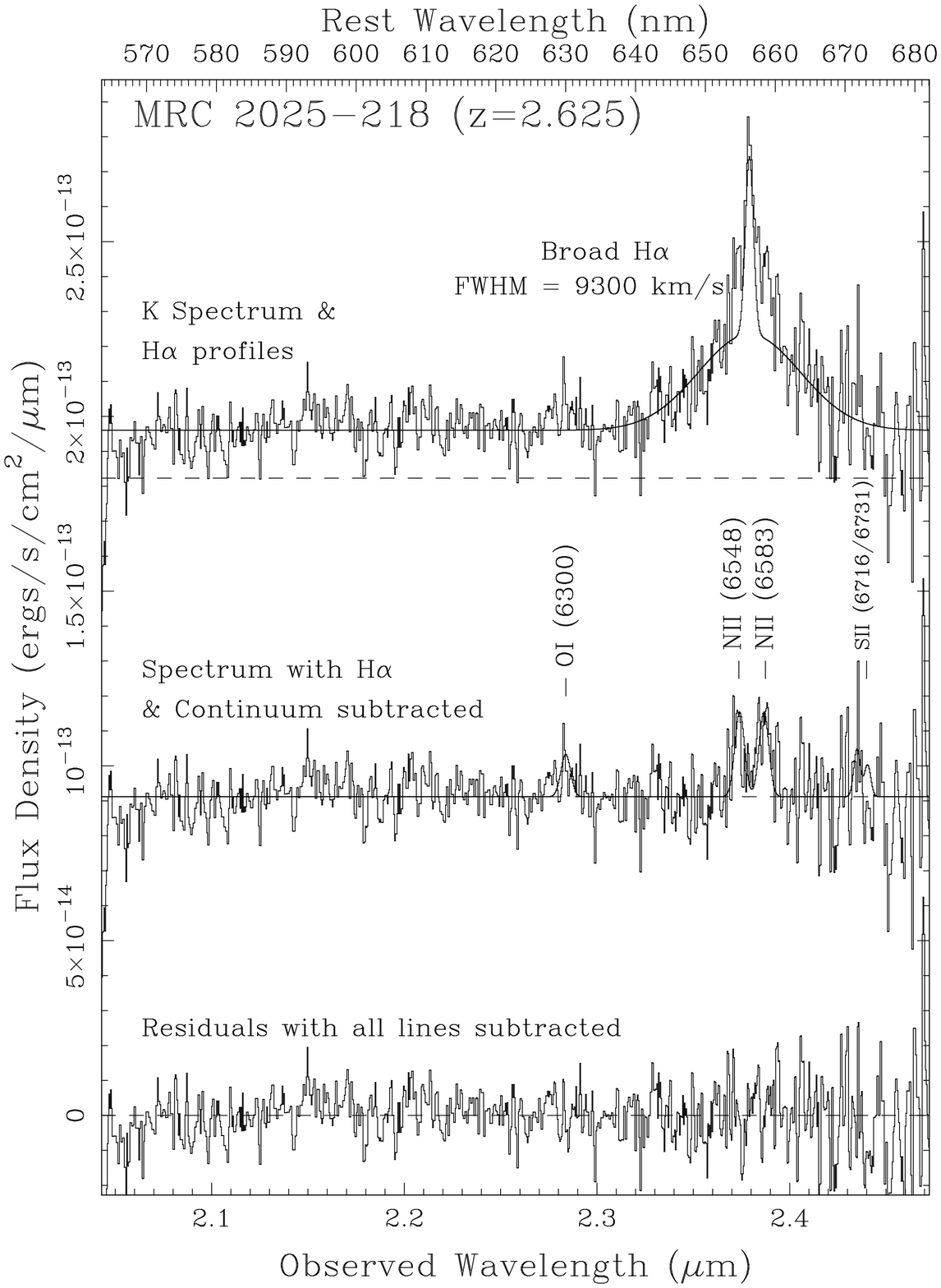}{4in}{0}{40}{40}{0}{0}}
{\footnotesize Figure 4. - The K band spectrum of MRC 2025-218 is
dominated by a very wide (9300 km s$^{-1}$) strong emission line of
H$\alpha$. The upper graph is the reduced spectrum overlayed with the
H$\alpha$ profiles.  The dashed line is the x-axis for this graph. The
middle graph has had the continuum and H$\alpha$ emission lines
subtracted to emphasize the weaker lines of NII.  The bottom graph is
the residuals after subtracting gaussians for each emission line.}
\medskip

Given the similarities in line width with radio loud quasars we now
try to determine if the extinction could explain the observed
differences between MRC~2025-218 and radio loud quasars.  If we
correct the H$\alpha$ flux for A$_V$=1.4 mag (the upper limit to the
narrow line extinction), then the broad line flux becomes
5.2$\times$10$^{-15}$ ergs~s$^{-1}$~cm$^{-2}$ or a broad line
H$\alpha$ luminosity of 2.6$\times$10$^{44}$ ergs s$^{-1}$.  We used
the sample of quasars of McIntosh et al. (1999) to derive a mean
H$\alpha$ luminosity of 6.11$\times$10$^{44}$ ergs s$^{-1}$ based on
the mean H$\beta$ equivalent width of their sample, no extinction and
an intrinsic ratio of 3.1 between H$\alpha$ and H$\beta$.  The one
sigma dispersion in this value is only 10\% in their sample.  Our
extinction corrected H$\alpha$ luminosity is then weaker than their
mean by a factor of 2.6 suggesting the central engines are very
similar. If we go a step further and assume that the intrinsic
luminosities are the same, then the broad line extinction would need
to be A$_V$=3.5 mag instead of A$_V$=1.4 mag as derived above for the
narrow line region.

A remaining difference between MRC~2025-218 and the quasars in the
McIntosh sample is the H-band magnitude. MRC~2025-218 has a broad band
magnitude of H=19.1 while the mean quasar H-band magnitude is
15.16. After correcting for the different redshifts (quasar mean
z=2.2) then MRC~2025-218 is 3.2 magnitudes fainter than the quasars at
a rest wavelength of 4550 \AA. If the broad band flux of MRC~2025-218
is dominated by the AGN then it would require A$_V$=4.2~mag to make it equal
to the quasar sample. This is surprisingly close to the value of 3.5 mag
required to match the broad H$\alpha$ fluxes. The
assumption that the AGN dominates the broad band flux in a radio
galaxy, however, is not obvious and may be in conflict with the
empirically determined K magnitude versus redshift relation observed
in both low redshift and high redshift objects (Eales et
al. 1997). MRC~2025-218 is consistent with the K vs. Z relation both
with and without taking the line emission into account.

Villar-Martin et al. (1999) find that MRC~2025-218 has large ratios
of [NV]/HeII and [NV]/[CIV] and suggest that the most likely explanation
is that N is overabundant.  They held out the possibility, however,
that contamination from a broad line region was enhancing this line
in comparison to their other radio galaxies.  But they argued against
this due to the lack of any broad lines including [CIII].  From our
broad H$\alpha$ detection, however, we clearly see that the broad line
region is only partially obscured and the strong NV emission is probably
not indicative of high metalicity.  This is further corroborated by
the relatively low ratios of [NII]/H$\alpha$(narrow).

\subsection{Spectral Shape and Extended Emission}

The double spectral peak found in [OIII] could be due to a high velocity
(200 km s$^{-1}$) cloud of gas or possibly a double active
nucleus. The unsmoothed H$\alpha$ narrow line is quite noisy but also
shows a double profile with a separation of 200 km s$^{-1}$.  Due to
the noise, however, we are not confident in the second H$\alpha$ peak.
If the second peak were due to a star forming region it would be
unlikely that the [OIII] line would be double as well since the
OIII/H$\beta$ ratio should be much lower for a starburst.

The off nucleus knots seen in [OIII] are difficult to
understand. Extended OIII has been observed in other radio galaxies
aligned to the radio axis (Armus et al. 1998) but no line ratios have
been determined for this gas.  If we assume that the emission is from
starbursts then our brightest knot (1$\times$10$^{-16}$ ergs s$^{-1}$
cm$^{-2}$) would have an [OIII]/H$\alpha$ ratio less than 1.0. This
would make the H$\alpha$ flux greater than
1$\times$10$^{-16}$~ergs~s$^{-1}$~cm$^{-2}$ and a luminosity more than
5$\times$10$^{42}$ ergs s$^{-1}$.  Assuming the relationship of
Kennicutt (1983) that the star formation rate is equal to L(H$\alpha$)
divided by 1.12$\times10^{41}$ ergs/s we derive a star formation rate
of 45 M$_{\odot}$ yr$^{-1}$.  This is comparable to the rates seen in
Pettini et al. (1998) where they studied 5 star forming galaxies in
the redshift range 2.2 to 3.3.  This is also close to the estimated
star formation rate of the Lyman Break Galaxy MS1512-cB58. As
calculated in Teplitz et al. (2000) cB58 has a SFR of 620 M$_{\odot}$
yr$^{-1}$ but after removing a factor of 30 for gravitational lensing
this becomes 21 M$_{\odot}$ yr$^{-1}$.

\section{Conclusions}

We have obtained the most sensitive infrared spectra ever taken of a
high redshift radio galaxy. The galaxy has very strong emission lines
with ratios and line widths consistent with an obscured quasar.  The
narrow line region appears to be partially obscured with A$_V$ around
1.4 mag, but from comparisons with high redshift quasars, we estimate
that the extinction to the broad line region is between 3 and 5
magnitudes.  Since other radio galaxies in the same redshift range
don't show broad emission lines, we suggest that MRC~2025-218 is
further along in its evolution towards an unobscured quasar.  We
cannot rule out any of the proposed mechanisms for the production of
the aligned emission. But based on the [OIII] line strength if the
majority of the emission is due to star formation, we find that the
star formation rate is comparable to that of Lyman Break Galaxies at
similar redshifts.  We urge even deeper observations of this and other
similar radio galaxies in order to measure additional extended line
emission.

\acknowledgments

It is a pleasure to acknowledge the hard work of past and present
members of the NIRSPEC instrument team at UCLA: Maryanne Angliongto,
Oddvar Bendiksen, George Brims, Leah Buchholz, John Canfield, Kim
Chin, Jonah Hare, Fred Lacayanga, Samuel B. Larson, Tim Liu, Nick
Magnone, Gunnar Skulason, Michael Specncer, Jason Weiss and Woon
Wong. In addition, we thank the Keck Director Fred Chaffee, CARA
instrument specialist Thomas A.  Bida, and all the CARA staff involved
in the commissioning of NIRSPEC. We also want to thank
Lee Armus for many useful discussions.  We are also grateful for a
very careful review from our anonymous referee. Data presented herein
were obtained at the W.M. Keck Observatory which was made possible by
the generous financial support of the W.M. Keck Foundation.


\begin{references}
\reference{asm} Antonucci, R. 1993, ARA\&A, 31, 473
\reference{asm} Armus, L., Soifer, B. T., Murphy, T. W., Neugebauer, G., Evans,
A. S., \& Matthews, K. 1998, \apj, 495, 276
\reference{bak} Baker, A. C., Carswell, R. F., Bailey, J. A., Espey, B. R.,
Smith, M. G., \& Ward, M. J. 1994, MNRAS, 270, 575
A. S., \& Matthews, K. 1998, \apj, 495, 276
\reference{bit} Binney, J. \& Tremaine, S. 1987, \it Galactic Dynamics \rm,
(Princeton: Princeton University Press
\reference{crd} Cardelli, J. A., Clayton, G. C. \& Mathis, J. S. 1989,
\apj, 345, 245
\reference{cmm} Chambers, K. C., Miley, G. K., van Breugel, W. J. M., Bremer,
M. A. R., Huang, J. S., \& Trentham, N. A. 1996, \apjs, 106, 215
\reference{cim} Cimatti, A., Dey, A., van Breugel, W., Antonucci, R.
\& Spinrad, H. 1996, \apj, 465, 145
\reference{deY} De Young, D. S. 1989, \apj, 342, L59
\reference{evans} Evans, A. S. 1998, \apj, 498, 553
\reference{fab} Fabian, A. 1991, MNRAS, 238, 41
\reference{grah} Graham, J. R., Carico, D. P., Matthews, K., Neugebauer, G.,
Soifer, B. T. \& Wilson, T. D. 1990, \apjl, 354, L5
\reference{ken} Kennicutt, R. 1983, \apj, 272, 54
\reference{mpw} McCarthy, P. J., Persson, S. E., \& West, S. C. 1992
\apj, 386, 52
\reference{mkv} McCarthy, P. J., Kapahi, V., van Breugel, W. \&
Subrahmanya, C. 1990, \aj, 100, 1014
\reference{mci} McIntosh, D. H., Rieke, M. J., Rix, H.-W., Foltz, C. B.
\& Weymann, R. J. 1999, \apj, 514, 40
\reference{mclean} McLean, I. S., et al. 1998, SPIE, 3354, 566
\reference{mclean} McLean, I. S., et al. 2000, PASP, in preparation
\reference{oster} Osterbrock, D.E. 1989, \it Astrophysics of Gaseous Nebulae
and Active Galactic Nuclei \rm (Mill Valley: University Science Books)
\reference{prm} Pentericci, L., Rottgering, H. J., A., Miley, G. K.,
McCarthy, P., Spinrad, H., van Breugel, W. J. M., \& Macchetto, F. 1999,
A\&A, 341, 329
\reference{pet} Pettini, M., Kellogg, M. Steidel, C. C., Dickinson, M.,
Adelberger, K. L. \& Giavalisco, M. 1998, \apj, 508, 539
\reference{ree} Rees, M. J. 1989, MNRAS, 239, 1
\reference{tep} Teplitz, H. I. 2000, \apjl, submitted
\reference{van} van Breugel, W. J. M., Stanford, S. A., Spinrad, H., Stern,
D., \& Graham, J. R. 1998, \apj, 502, 614
\reference{vima} Villar-Martin, M., Binette, L. \& Fosbury, R. A. E.
1999a, 346, 7
\reference{vim} Villar-Martin, M., Fosbury, R. A. E., Binette, L., 
Tadhunter, C. N., \& Rocca-Volmerange, B. 1999b, accepted for A\&A
\end{references}
\end{document}